\documentclass{PoS}

\title{Bounds on Plack-scale deformation of CPT from lifetimes and interference}

\ShortTitle{CPT Planck-scale deformation}

\author{\speaker{Wojciech Wi\'slicki}\thanks{This research was supported by the Polish National Science Centre grant nr 2017/26/M/ST2/00697 and by the Munich Institute for Astro- and Particle Physics (MIAPP) of the DFG Excellence Cluster Origins (www.origins-cluster.de).}\\
        National Centre for Nuclear Research, Warsaw, Poland\\
        E-mail: \email{wojciech.wislicki@ncbj.gov.pl}}

\abstract{
Deformed relativistic kinematics, expected to emerge in a flat-spacetime limit of quantum gravity, predicts violation of discrete symmetries at energy scale in the vicinity of the Planck mass.
Momentum-dependent deformations of the C, P and T invariance are derived from the $\kappa$-deformed Poincar\'e algebra.
Deformation of the CPT symmetry leads to a subtle violation of
Lorentz symmetry.
This entails some small but measurable phenomenological consequences, as corrections to characteristics of time evolution: particle lifetimes or frequency of flavour oscillations in two-particle states at high energy.
We argue here that using current experimental precisions on the muon lifetime one can bound the deformation parameter $\kappa>10^{14}$ GeV at LHC energy and move this limit even to $10^{16}$ GeV at Future Circular Collider, planned at CERN.
Weaker limits on deformation can be also obtained from interference of neutral mesons. In case of ${\mathrm B^0}$s from $\Upsilon$ decay it amounts to $\kappa> 10^8$ GeV at confidence level 99\%.
}

\FullConference{
European Physical Society Conference on High Energy Physics - EPS-HEP2019 -\\
			10-17 July, 2019\\
			Ghent, Belgium}

\begin{document}

\section{Introduction}
Invariance under CPT, the combined transformation of the space inversion P, charge conjugation C and time reversal T, is believed to be strictly obeyed due to theorems based on premises constituting natural axioms of the local quantum field theory \cite{streater}.
Among phenomenological consequences of the CPT theorem are equality of masses and lifetimes of particles and antiparticles.
These claims were experimentally tested with high accuracy. The most precise constraint comes from the neutral kaon physics where masses of the $\mathrm{K^0}$ and $\mathrm{\bar K^0}$ are equal with accuracy of $4\times 10^{-19}$ GeV at 95 \% confidence level \cite{pdg} \footnote{The number given in ref. \cite{pdg} represents the world average where the main contribution comes from the CPLEAR experiment at CERN \cite{cplear_long}.}. 
Together with the anti-CPT theorem \cite{greenberg} (cf. discussion in ref. \cite{chaichian}), stating that CPT violation entails violation of the Lorentz invariance, these theoretical results may suggest that any experimental hint of CPT non-conservation might deeper affect the quantum field theory.  

Effects of quantum gravity have long been suggested as a possible source of CPT violation.
As noted in ref. \cite{wald}, there is no fundamental arrow of time in its own right but only associated with a choice of matter or antimatter.
In addition, in presence of an inherent quantum-gravitational background, CPT operator is no longer well defined. Scattering operator cannot map pure {\it in}- into {\it out}-states, and {\it vice versa}, due to destruction of information in presence of micro black holes \cite{hawking}. 
Because of that, any system propagating in a quantum-gravitational background exhibits irreversibility, analogously to dissipative processes but here connected to CPT violation.
It has therefore inspired approaches based on dissipative quantum dynamics (for historical account cf. ref. \cite{kossakowski}) and resulting with a number of models with the CPT- or Lorentz violation ascribed to gavity-induced quantum decoherence \cite{decoherence}.
Alternatively, another approach inspired by quantum gravity was considered \cite{kostelecky} using the usual framework of quantum field theory but with an explicitly CPT non-invariant term added to the Lagrangian.
These models, either based on decoherence or Standard Model extensions, were considerably, but not definitely, constrained by experimental data on neutral kaons \cite{kloe1,kloe2}.

\section{Planck-scale Deformation and Discrete Symmetries}

By analogy to the Heisenberg uncertainty relation, the concept of minimal energy and length scales leads to non-commutative geometry, defined by the commutation relations
\begin{eqnarray} \label{eq1}
[x^\mu,x^\nu]=i\theta^{\mu\nu},\quad\quad \mu,\nu=0,1,2,3 
\end{eqnarray} 
where $\theta^{\mu\nu}$ can be related to geometric properties of the space-time \cite{hossenfelder}.
The concept has been further developed to the geometry generated by the $\kappa$-deformed Poincar\'e algebra - generating Lorentz boosts, momenta and rotations - and defined by the commutation relations
\begin{eqnarray} \label{eq2}
[t,x^j]=ix^j/\kappa,\quad\quad [t,k^j]=-ik^j/\kappa,\quad\quad j=1,2,3
\end{eqnarray}
where $\kappa$ is expected to be of the order of Planck's mass $m_P\simeq 2\times 10^{18}$ GeV$/c^2$ \cite{kowalski1}. 
The $\kappa$-deformed momentum space is a submanifold of the four-dimensional de Sitter space, defined in the five-dimensional Minkowski space by the constraint
\begin{eqnarray}\label{eq5}
-p_0^2+p_1^2+p_2^2+p_3^2+p_4^2=\kappa^2;\quad\quad p_0+p_4>0,
\end{eqnarray} 
where $\kappa$ is related to the curvature of the momentum manifold.

Essential to understand the action of discrete symmetries in deformed space is a proper definition of deformed rule of the four-momentum composition and thus of momentum inversion.
As elaborated in ref. \cite{arzano1} and using the Hopf algebra, the inverse of three-momentum is given by its {\it antipode} $\ominus$ as
\begin{eqnarray}\label{eq6}
S(p_i)\equiv \ominus p_i= -p_i\frac{\kappa}{E+p_4}, \quad\quad i=1,2,3
\end{eqnarray}
and for energy by requiring the mass-shell relation or preservation of the Casimir operator
\begin{eqnarray}\label{eq7}
m^2 & = & E^2 - {\bf p}^2 \nonumber \\
    & = & S(E)^2 - S({\bf p})^2,
\end{eqnarray}
by the formula
\begin{eqnarray}\label{eq8}
S(E) = \frac{\kappa^2}{E+p_4}-p_4.
\end{eqnarray}
Using these findings, action of the $\kappa$-deformed CPT transformation, denoted henceforth as $\Theta_\kappa=$CPT$_\kappa$, on the four-momentum can be written in leading order of $1/\kappa$ as
\begin{eqnarray}\label{eq9}
\Theta_\kappa p_0 & = & p_0-\frac{{\bf p}^2}{\kappa} + {\mathcal O}(1/\kappa^2) \nonumber \\
\Theta_\kappa {\bf p} & = & {\bf p}-\frac{p_0{\bf p}}{\kappa} + {\mathcal O}(1/\kappa^2)
\end{eqnarray}
($p_0=E$) and charges are always multiplied by $-1$. 
In particular, the particle's Lorentz boost factor $\gamma=E/m$, after $\Theta_\kappa$ deformation becomes
\begin{eqnarray}\label{eq10}
\gamma_\kappa=\frac{1}{m}(E-{\bf p}^2/\kappa)
\end{eqnarray}
for the antiparticle.

\section{Measurement of the $\kappa$-Deformation}

Consider unstable particle at rest described by the wave funcion depending on its proper time
\begin{eqnarray}\label{eq11}
\psi=\sqrt{\Gamma}e^{-\Gamma t/2+imt},
\end{eqnarray}
where its mass $m$ and lifetime $\tau=1/\Gamma$ are CPT-invariant.
In the particle's rest frame the CPT is undeformed and the particle's ($_p$) and antiparticle's ($_a$) masses and lifetimes are equal due to the CPT theorem.
Their decay probabilities obey the same decay laws
\begin{eqnarray}\label{eq12}
{\mathcal P}_p={\mathcal P}_a=\Gamma e^{-\Gamma t}
\end{eqnarray}
but they differ after Lorentz transformation to moving system
\begin{eqnarray}\label{eq13}
{\mathcal P}_p & = & \frac{\Gamma E}{m} e^{-\Gamma E/m t} \nonumber \\
{\mathcal P}_a & = & \Gamma(\frac{ E}{m}-\frac{{\bf p}^2}{\kappa m}) e^{-\Gamma (\frac{ E}{m}-\frac{{\bf p}^2}{\kappa m}) t}.
\end{eqnarray}
Consequences of deformation could thus be examined experimentally by precisely measuring the particle and antiparticle lifetimes.
These are equally delated due to Lorentz boost but only one of them is deformed under CPT transformation. 
To be measurable, effective correction ${\bf p}^2/(\kappa m)$ has to be comparable to experimental accuracy of the measurement $\sigma_\tau/\tau$. 

In the scheme described above and developed in ref. \cite{wislicki}, violation of the CPT is momentum-dependent and thus depends on the Lorentz frame. 
It thus explicitly relates the CPT- and Lorentz noninvariance, as suggested by the CPT theorems in general terms.

For known unstable particles the accuracy of measurements of their lifetimes amounts typically between $10^{-4}$ for the mesons like ${\mathrm \pi^\pm}$ and ${\mathrm K^0}$ and $10^{-6}$ for leptons $\mu^\pm$ \cite{pdg}. 
Any measurement of the lifetime requires experimentally measured momenta to be Lorentz-transformed to the particle$'$s rest frame.
Inaccuracies of laboratory momenta and energies thus propagate to the rest frame and, if large, can strongly affect $\sigma_\tau$. 
This is particularly discouraging in non-accelerator experiments where energies of cosmic particles are occasionally very high, exceeding even $10^6$ GeV and thus ${\mathbf p}^2 \sim 10^{12}$ GeV$^2$ but, at the same time, experimental uncertainties
usually tend to be large and hard to control.

In order to quantify our findings, in Fig. 1 (upper left) we plot the correction ${\mathbf p}^2/(\kappa m)$ for the muon.
If deformation $\kappa$ is close to the Planck mass $10^{19}$ GeV, as expected, any detectable correction requires momenta of the order $10^6$ GeV, unattainable at today$'$s accelerating facilities. 
Such energies are available in cosmic-ray experiments. 
However, using them for our purposes would require a measurement of their lifetimes and reach very challenging accuracy of their energy determination.
We can also estimate a limit on the deformation $\kappa$ that can be set for present energies at LHC and those planned at FCC \cite{fcc}, both at CERN. 
Using experimental accuracies of the lifetimes and requiring ${\mathbf p}^2/( \kappa m ) = \sigma_\tau$ for $p = 6.5$ TeV (LHC) and 50 TeV (FCC) one obtains the values of $\kappa$ labeling curves in Fig. 1. 
As can be seen there, the limiting value of $\kappa=4\times 10^{14}$ GeV can be obtained using muons at LHC and, in future, $\kappa=2\times 10^{16}$ GeV at FCC. 
Further improvement at these energies requires progress in accuracy of the lifetime measurement.
\begin{figure}[htb]
\begin{tabular}{cc}
\hspace{1cm} \includegraphics[width=5cm]{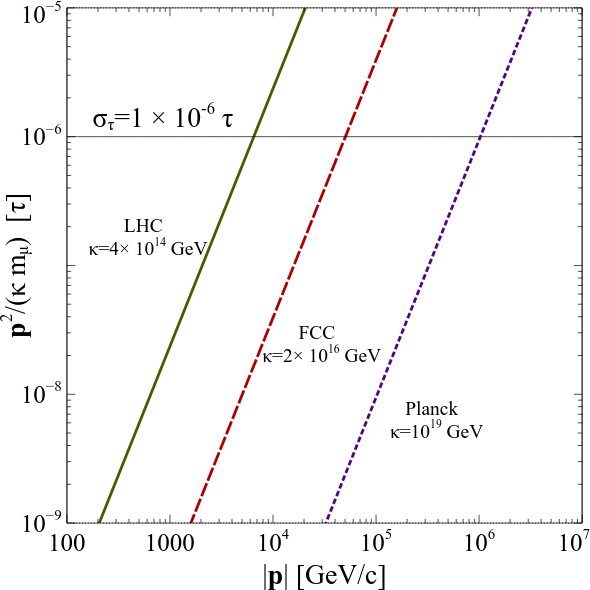} & \hspace{1cm} \includegraphics[width=6cm]{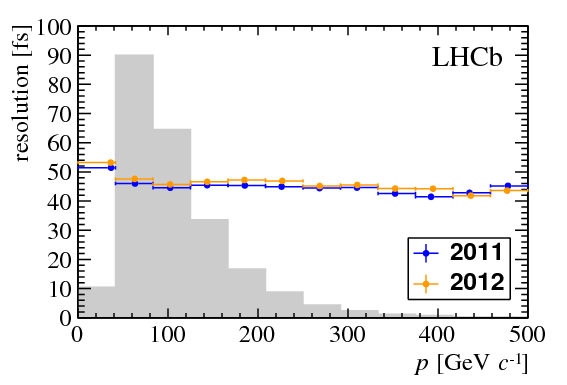} \\
\hspace{1cm} \includegraphics[width=5cm]{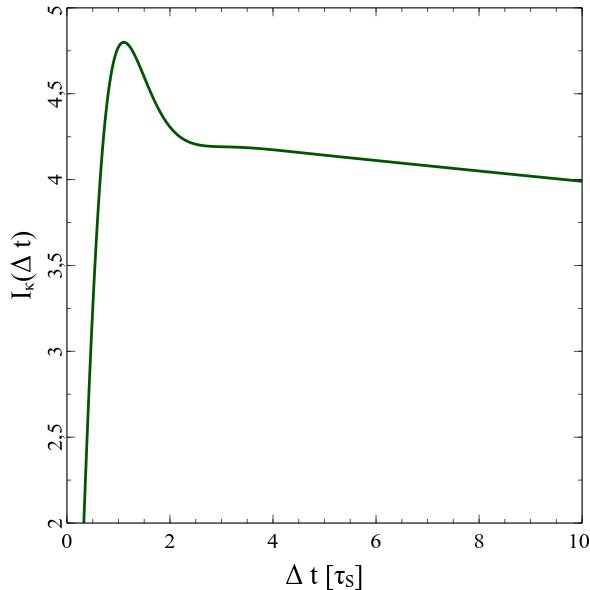} & \hspace{1cm} \includegraphics[width=5cm]{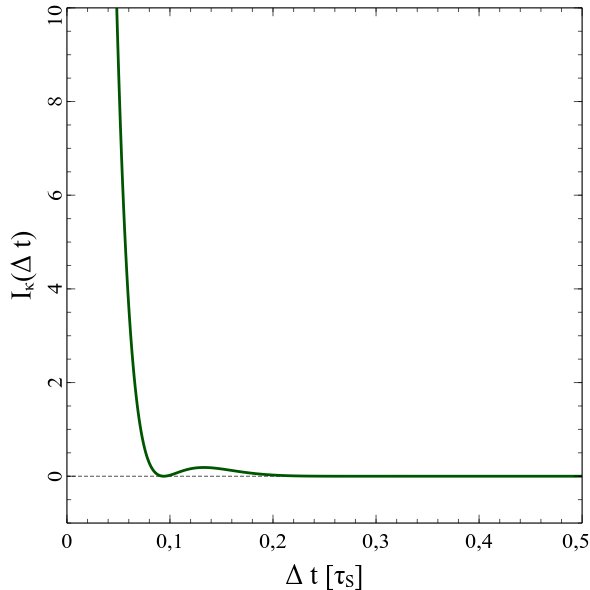}
\end{tabular}
\caption{{\em {\bf Upper left}: Correction ${\mathbf p}^2/(\kappa m)$ to the muon lifetime.  Two curves, labeled LHC and
FCC, are for the deformation parameters $\kappa$ corresponding to corrections equal to
experimental accuracies for maximal momenta at the Large Hadron Collider (continuous green, LHC) and the Future Circular Collider (dashed red, FCC). The violet dotted line corresponds to the Planck mass $\kappa=10^{19}$ GeV. {\bf Upper right}: Momentum-dependent time resolution of the LHCb detector. {\bf Lower left}: Lorentz-boosted spectra $\gamma=4.3$ of decay time difference for pairs of kaons from decay ${\mathrm \phi(1020)\rightarrow K_LK_S}$, in units of $\tau_S$. {\bf Lower right}: Lorentz-boosted spectra, $\gamma=44$ of decay time difference for pairs of kaons from decay ${\mathrm \Upsilon(10580)\rightarrow B_H B_L}$, in units of $\tau_L$.}}
\label{Fig:F2H}
\end{figure}

Interesting possibilities of search for CPT violation are given by interferometry of neutral mesons produced in coherent, two-particle states in decays of pseudoscalar mesons, \linebreak e.g. ${\mathrm \phi(1020)\rightarrow K_LK_S}$ and ${\mathrm \Upsilon(10580)\rightarrow B_H B_L}$ (cf. refs \cite{decoherence}).
Interference patterns are usually studied in difference of decay times, $\Delta t=|t_2 - t_1|$
\begin{eqnarray}\label{eq14}
I(\Delta t)\sim e^{-\Gamma_L \,\Delta t}+e^{-\Gamma_S \,\Delta t}-2\, e^{-\bar\Gamma\,\Delta t}\cos(\Delta m\,\Delta t),
\end{eqnarray}
where $t_{1(2)}$ stands for decay time of the first (second) meson in its rest frame and $\Delta m=m_L-m_S$ for difference of masses of the heavier and lighter of two neutral mesons and $\bar\Gamma=(\Gamma_L+\Gamma_S)/2$.
Applying deformed CPT (\ref{eq7}) transforms decay spectrum (\ref{eq14}) to
\begin{eqnarray}\label{eq15}
I(\Delta t) & \sim & (\gamma-{\mathbf p}^2/(m\kappa))(e^{-\gamma\,\Gamma_L\,\Delta t}+e^{-\gamma\,\Gamma_S\,\Delta t}) \nonumber \\
            & +    & \gamma\,\Delta t\,{\mathbf p}^2/(m\kappa)(\Gamma_L e^{-\gamma\,\Gamma_L\,\Delta t}+\Gamma_S e^{-\gamma\,\Gamma_S\,\Delta t}) \nonumber \\
            & -    & 2\gamma \,e^{-\gamma\,\bar\Gamma\,\Delta t}\,[(1+\bar\Gamma\,\Delta t\,{\mathbf p}^2/(m\,\kappa))\cos(\gamma\,\Delta\, m\Delta t) \nonumber \\
            & + & \Delta m\,\Delta t\,{\mathbf p}^2/(m\,\kappa)\sin(\gamma\,\Delta m\,\Delta t)],
\end{eqnarray}
where $m=(m_L+m_S)/2$.
The Lorentz boost, by increasing masses of mesons, effectively amplifies oscillation frequency $\gamma\,\Delta m$  of the oscillatory terms .
In order to get the interference pattern experimentally measurable, the amplified frequency cannot exceed inverse time resolution of an apparatus $\frac{1}{\gamma \Delta m} > \sigma_t$.
Time resolution of the most precise spectrometer at LHC amounts to 45 fs and belongs to the LHCb spectrometer (cf. Fig.1, upper right). 
It gives constraints to the maximum Lorentz boost for pairs of ${\mathrm K^0}$ and ${\mathrm B^0}$ mesons to be $\gamma=4.3$ and $\gamma=44$, respectively.
Decay time spectra (\ref{eq15}) for these boosts are presented in Fig. 1 (lower left and right).
Limits on $\kappa$ deformation can be estimated by Monte Carlo using the log-likelihood method and are found to be $2\times 10^5$ GeV and $1.2\times 10^8$ GeV, at 99.9\% confidence level, for pairs of ${\mathrm K^0}$s and ${\mathrm B^0}$s, respectively.

\section{Conclusions}

Deformed CPT transformation can be used to estimate the deformation parameter $\kappa$ from experiment. 
The CPT-violating corrections to energy-momentum are of the order ${\mathbf p}^2/(m\kappa)$.
This kind of CPT violation automatically violates Lorentz invariance.
Numerical estimates show that using precisely known lifetimes of $\mu^\pm$ one could expect to limit $\kappa$ at $10^{16}$ GeV, i.e. only three orders of magnitude lower than the Planck mass, by incorporating the FCC at $\sqrt{s}=100$ TeV. 
Limitations from neutral meson interferometry are less stringent.

\end{document}